# Thermal Dynamics of Graphene Edges Investigated by Polarized Raman Spectroscopy

Ya Nan Xu[†,‡,⊥], Da Zhan[†,⊥], Lei Liu[†], Hui Suo[‡], Zhen Hua Ni[§,*], Thuong Thuong Nguyen[∥], Chun Zhao[‡], and Ze Xiang Shen[†,*]

[†] Division of Physics and Applied Physics, School of Physical and Mathematical Sciences, Nanyang Technological University, 637371, Singapore.
[‡] State Key Laboratory on Integrated Optoelectronics, College of Electronic Science and Engineering, Jilin University, 2699 Qianjin Street, Changchun 130012, China.
[§] Department of Physics, Southeast University, Nanjing, 211189, China
[∥] Department of Physics and Astronomy, University of Leeds, Leeds LS2 9JT, UK.
* Address correspondence to zhni@seu.edu.cn; zexiang@ntu.edu.sg
[⊥] These authors contributed equally to this work.

**ABSTRACT**

In this report, we present Raman spectroscopy investigation of the thermal stability and dynamics of graphene edges. It was found that graphene edges (both armchair and zigzag) are not stable and undergo modifications even at temperature as low as 200°C. Based on polarized Raman results, we provide possible structural models on how graphene edges change during annealing. The zigzag edges rearrange and form armchair segments that are ±30° relative to the edge direction, while armchair edges are dominated by armchair segments even at annealing temperature as high as 500°C. The modifications of edge structures by thermal annealing (zigzag segments rearrange in form of armchair segments) provide a flexible way to control the electronic properties of graphene and graphene nanostructures.

Graphene attracted great attention since it was first realized in 2004.[1-5] The unique electronic properties make graphene a promising candidate for ultrahigh-speed nanoelectronics,[1-8] particularly graphene nanoribbons (GNRs) have been studied extensively because it was predicted that its electronic properties can be modulated by variation of its edge properties.[9-13] There are two standard types of edges: the zigzag edges (Z-edges) and armchair edges (A-edges) (Figure 1a and 1b). Theoretical calculations predicted that A-edges GNRs are semiconducting, with an energy band gap inversely proportional to the ribbon width,[9] while Z-edges GNRs are metallic.[14] However, Son *et al*. suggested that the band gap can also be observed even for Z-edges GNRs considering spin effect.[10,15] Recently, Ritter and Lyding experimentally demonstrated that predominately Z-edges GNRs also exhibit a finite energy gap, but it significantly decreases with higher concentration of zigzag segments.[13] Therefore, tailoring the electrical/electronic properties of GNRs by modifying its edge chirality or shape, for example by thermal annealing, would be a very interesting research topic. Furthermore, graphene edges are preferred sites for functionalization and chemical decoration,[16] while thermal annealing is commonly used for sample cleaning[17-19] as well as in semiconductor manufacturing process.[2,7,8,11,13,17,19-21] Therefore, a good understanding of the thermal stability and dynamics of graphene edges is important for guiding future graphene-based nanoelectronics processing and chemical decoration. Recently, Girit *et al*. probed the edge stability and dynamics by using high resolution TEM.[12] However, the energy of electron beam used in TEM is very large and the temperature induced might be very high. Therefore, the *in situ* TEM study of graphene edge dynamics is actually the study of stability of graphene edges against electron collision or irradiation, and it is difficult to control the experimental conditions, *i.e.* effective temperature induced by irradiation.

Raman spectroscopy is one of the most commonly used methods for characterization of graphene-based materials because of the advantages of simple, non-destructive and high-throughput. It is also proved as one of the most effective methods for identifying types of graphene edges.[22] The Raman D mode (Breathing mode $A_{1g}$) is inactive for perfect Z-edges, because the exchanged momentum by scattering from the Z-edges ($d_Z$) cannot connect the adjacent Dirac cones K and K' (Inset of Figure 1a), hence does not fulfill the double resonance process.[23-25] On the other hand, the exchanged momentum from A-edges can satisfy the intervalley scattering process between K and adjacent K' (Inset of Figure 1b); hence its D band is Raman active. Furthermore, the perfect A-edges gives rise to the polarization dependence of the D peak intensity as $I(D) \propto \cos^2\theta$, where $\theta$ is the angle between incident laser polarization and the edge direction throughout this paper. In this paper, the thermal stability and dynamics of graphene edges for both armchair and zigzag are systematically investigated by polarized Raman spectroscopy. It was found that edges of graphene are not stable and rearrange even at 200°C. Structure models for annealed graphene edges are proposed based on the polarization Raman results.

**RESULTS AND DISCUSSION**

Figure 2a is the optical image of a single layer graphene flake (sample a) with angle between adjacent edges of 150°. Figure 2b and 2c are Raman images of intensities of G and D peaks of this sample. It can be seen clearly from Figure 2c that for the pristine graphene, edge $a_1$ presents a strong D peak while edge $a_2$ shows only a very weak D peak. To avoid the Raman intensity dependence on the relative angle between the edge and laser polarization direction, the incident laser is polarized (shown in Figure 2b) along the direction of the bisector. Therefore, it can be confirmed that edge $b_1$ is A-edge while edge $b_2$ is Z-edge.[22] Following, the sample is annealed in vacuum ($5\times10^{-5}$ mbar) in order to study the thermal stability of graphene edges. Figure 2d presents Raman image of D peak intensity of the same sample after annealed at 300°C for 15 minutes. It can be clearly seen that the D peak intensity of edge $a_2$ becomes comparable to that of edge $a_1$. This suggests that there is a significant change for the Z-edges (edge $a_2$) after thermal treatment, where the edge atoms may undergo rearrangement to satisfy double resonance and hence activate the Raman D peak. Raman spectra collected from the edges (Figure 2e) also demonstrate the obvious increase of the D peak intensity of Z-edge $a_2$ after annealing. In the meanwhile, the Raman spectra from the center region of graphene flake do not show any D peak (not shown). For the other three samples which were annealed at 200, 400 and 500°C (sample b, c and d), the Raman images show similar changes, with a significant increase of D peak intensity of Z-edges after annealing (as shown in Figure S1- S3 of supplementary information). However, the intensity of D peak after 200°C annealing is relatively small, which suggest that 200°C is the critical temperature where Z-edges start to change/rearrange. Our results suggest that graphene edges (A-edges to be discussed later) are not stable even at temperature as low as ~200°C. Such annealing temperature (or even higher temperature) is commonly used for graphene cleaning as well as process in device fabrications.[2,7,8,11,13,17,19-21] Furthermore, we should point out that because vacuum annealing can naturally induce p-doping on graphene.[26,27] This would definitely give rise to the stiffening (blueshift) and sharpening effect on G peak as shown in Figure 2e, due to the non-adiabatic removal of the Kohn anomaly at Γ point.[28,29] For the 2D peak of annealed sample, the significantly decrease of intensity compare to that of pristine graphene is observed (not shown), which is in agreement with the previous reported result elsewhere;[30] and this is also because that doping gives rise to increase of the electron-electron collision and inelastic scattering rate which strongly decreased the 2D peak intensity.[31,32]

As aforementioned that Z-edges do not fulfill double resonance condition and hence do not present D peak. The appearance of small D peak in pristine Z-edges (<5% of the intensity of G peak as shown later in Figure 5a) is attributed to small amount of armchair segments and short range defects.[22,33] For the significant increase of D peak intensity of annealed Z-edges, two possible edge modifications are proposed: (1) carbon atoms of Z-edges reconstructed to form armchair segments ±30° respect to the edge direction (A-30°), as shown in Figure 3b and 4a; (2) short-range defects formed, *i.e.* the edge structures with coherence length less than the electron wavelength 0.6 nm (for electron excited by 532 nm laser)[33] cannot be considered as proper edges, instead they can be considered as point defects. The possible formation of C-O bonds at graphene edges during annealing can also be considered as short-range/point defects.[34] Note that the 5-7-7-5 and 5-8-5 defects are excluded in the samples as these defects normally formed at much higher temperature.[35,36] Besides the two proposed edge segments, some of the zigzag segments may remain unchanged (Z-0°), while some may rearrange as zigzag segments at ±60° respect to the original zigzag direction (Z-60°) during the annealing process, although these two components are D peak inactive. The structure of ideal pristine Z-edges and rearranged structure of annealed Z-edges are shown in Figure 3a and 3b, respectively.

In order to estimate the portions of each type of edge segments in annealed edges, polarized Raman studies were carried out. As different types of edge structures scatter electrons in different directions, the D peak induced by them have different polarization dependence, *i.e.* the dependence of D peak intensity of ideal A-edges on incident laser polarization can be expressed as $I_D \propto \cos^2\theta$. On the other hand, for short-range/point defects, the D peak intensity does not have any polarization dependence.[33,37] Therefore, the polarization studies of graphene edges can be used as a useful way to estimate the amount of each type of edge segments. For annealed Z-edge, both rearranged A-30° edges and point defects are D peak activated, while only A-30° defects present incident laser polarization dependence. The detailed structure of rearranged A-30° defects is shown in Figure 4a, the dependence of D peak intensity of A-30° edges on incident laser polarization can be expressed as:

$$I_D \propto \frac{1}{2}[\cos^2(\theta-30)+\cos^2(\theta+30)] \quad (1)$$

For comparison, the polarization dependence of D peak intensity for ideal A-edges (i.e. $\cos^2\theta$), A-30° edges ($\frac{1}{2}[\cos^2(\theta-30)+\cos^2(\theta+30)]$), and point defects are drawn in

Figure 4b. It is obvious that the D peak from A-30° edge has weaker polarization dependence compared with that from an ideal A-edge, while that from point defects does not show any polarization dependence. If we assume that in annealed Z-edges, the A-30° edges and point defects weight $f_1$ and $f_2$, respectively; while unchanged zigzag segments Z-0° and those rearranged Z-60° weight $f_3$ and $f_4$. Thus $f_1 + f_2 + f_3 + f_4 = 1$, and the polarization dependence of D peak intensity can then be expressed by:

$$\frac{f_1}{2}[\cos^2(\theta - 30) + \cos^2(\theta + 30)] + f_2 \quad (2a)$$

$$\text{or} \quad \frac{1}{4}f_1 + f_2 + \frac{1}{2}f_1 \cos^2\theta \quad (2b)$$

Figure 5a to 5e show the experimentally obtained $I_D/I_G$ as function of θ for pristine Z-edges and Z-edges annealed at 200°C, 300°C, 400°C and 500°C, respectively (In this paper, we use the intensity ratio of $I_D/I_G$ instead of absolute values of D peak intensity in order to minimize measurement error). The D peak of pristine Z-edges (Figure 5a) is very weak but non-zero as edges are not perfect in real case. For 200°C annealed Z-edges (Figure 5b), the $I_D/I_G$ becomes obviously larger than that of pristine Z-edges. The polarization dependence of $I_D/I_G$ presents similar shape as that expected from ideal A-30° edges (the red curves in Figure 4b and 5b). Thus it can be concluded that a small portion of zigzag segments in Z-edges start to modify and form A-30° segments at the annealing temperature of 200°C. For Z-edges annealed at higher temperature (300°C or above), the D peak intensity grows much stronger and is comparable to that of annealed A-edges (as shown in Figure 6a). This suggests that the values of $f_3$ and $f_4$ are very small compared with $f_1 + f_2$ (or else, the appearance of zigzag segments would certainly reduce the intensity of D peak), and could be ignored. Fitting the data in Figure 5b, 5c, 5d and 5e with formula (2), we can roughly obtain the values of $f_1/f_2$, which are ~80, ~14.5, ~6.6 and ~4.9 for Z-edges annealed at 200°C, 300°C, 400°C and 500°C, respectively. The portion of armchair components (A-30°) on Z-edges annealed at different temperatures is shown by the red triangles in Figure 6b. (The D peak intensity of 200°C annealed Z-edges is only ~35% of that of pristine A-edges, taking into account the shape of A-30°, the portion of armchair segments (A-30°) for 200°C annealed sample is estimated to be ~54%.) It can be seen that, the portion of A-30° segments increases dramatically at ~200°C (~54%) and reaches maximum at ~300°C (~93%). At higher temperature (400 and 500°C), A-30° segments still dominate in the annealed Z-edges (87% and 83% respectively), but with the portion of short-range point defects slightly increases.

Different from Z-edge, the D peak intensities of A-edges do not have obvious change after annealing at different temperatures except that the polarization dependence of D peak intensity becomes much weaker after annealing. The intensity ratio of $I_D/I_G$ of different A-edges and Z-edges with laser polarization at 90° and 0° relative to edges are shown in Figure 6a. It can be seen that the ratio between maximum $I_D/I_G$ ($\theta=0°$) and minimum $I_D/I_G$ ($\theta=90°$) of pristine A-edges is ~7 before annealing, while after annealed at 200, 300, 400 and 500°C, this ratio decreased to ~3.5, ~3.0, ~2.3 and ~2.3, respectively. These results clearly indicate that A-edges are also not stable and undergone modifications during annealing. The schematic crystal structures of pristine and annealed A-edges are shown in Figure 3c and 3d. For A-edge, we assume the armchair segments remain unchanged (A-0°) weights $f_5$; those rearranged armchair segment at ±60° (A-60°) and zigzag segments at ±30° (Z-30°) weight $f_6$ and $f_7$, respectively; the point defects weights $f_8$. Therefore, $f_5+ f_6+ f_7+ f_8=1$. Giving the factor that $(I_D/I_G)_{\theta=0°}+(I_D/I_G)_{\theta=90°}$ of annealed A-edges kept almost unchanged from pristine A-edges (Figure 6a), it would be reasonable to claim that Z-30° segments ($f_7$) are negligible (or else, the appearance of zigzag segments would certainly reduce the intensity of D peak). The polarization dependence of D peak intensity can then be written as:

$$f_5 \cos^2\theta + \frac{f_6}{2}[\cos^2(\theta - 60) + \cos^2(\theta + 60)] + f_8 \quad (3a)$$

$$\text{or} \quad \frac{1}{2}(2f_5 - f_6)\cos^2\theta + \frac{3}{4}f_6 + f_8 \quad (3b)$$

Figure 5f shows the $I_D/I_G$ as function of laser polarization for pristine A-edge. As can be seen, for pristine A-edge, $I_D/I_G$ almost follows $\cos^2\theta$ polarization dependence, and the value of $(I_D/I_G)_{\theta=0°}/(I_D/I_G)_{\theta=90°}$ is very large, this suggests that A-0° segments ($f_5$) dominate in A-edges. Fit the data in Figure 5f and data from other three pristine samples with formula (3), it can be obtained that the portion of A-0° ($f_5$) segments is around ~83% for pristine A-edges, and the remaining ~17% could mostly be A-60° segments as portion of point defects are very small at low temperature according to the results of annealed Z-edges. After annealing, the polarization dependence of $I_D/I_G$ becomes much weaker, as shown in Figure 5g to 5j. The portion values of A-0° obtained from fitting as a function of annealing temperature is shown in Figure 6b. It is clearly seen that the portion of A-0° becomes less with higher temperature annealing, but still values ~57% even after 500°C anneal. The rest edge components consist of A-60° and point defects, and we are not able to extract their exact portions based on formula (3) only. However, the ratio between A-60° and point defects can be roughly estimated from the results of annealed Z-edges, which are ~80, ~14.5, ~6.6 and ~4.9 for edges annealed at 200°C, 300°C, 400°C and 500°C, respectively. With such estimation, the portion of A-0°, A-60°, and point defects of annealed A-edges can be calculated respectively. Here, we mainly concern the armchair segments (A-0°+A-60°) for annealed A-edges, which are shown by blue stars in Figure 6b. It can be seen that, the armchair segments (A-0°+A-60°) dominate on annealed A-edges at different temperatures, with a slight drop at high temperature (300-500°C) due to the appearance of point defects.

Based on above discussion, we could conclude that armchair segments are more stable compared to zigzag segments during thermal treatment, *i.e.* zigzag would change and form armchair at high temperature. This agrees well with the theoretical calculation that armchair have lower energy compared to zigzag.[38] However, this result is different from the stability of graphene edges under electron beam irradiation, where the zigzag edge is relatively more stable than armchair.[12] This is possible due to the very different experimental conditions for the two experiments. There was also argued that the "zigzag" segments observed in that experiment are actually "reconstructed zigzag" (5-7 defects).[36] Meanwhile, we notice that graphene center region is very stable as the D-peak keeps inactive even we annealed graphene at 700°C ($5 \times 10^{-5}$ mbar).

It is theoretically predicted that A-edges GNRs are semiconducting with energy gap inversely proportional to the width,[9] and it is also experimentally demonstrated that the higher concentration of zigzag segments would significantly reduce the energy gap of GNRs.[13] Thus, the modifications of edge structures by thermal annealing (zigzag segments rearrange in form of armchair segments) provide a flexible way to control the electronic structure (such as energy gap) of graphene and graphene nanostructures.

**CONCLUSIONS**

In summary, it is experimentally confirmed by polarized Raman spectroscopy that different from center region of graphene, the graphene edges are not thermally stable even at low annealing temperature, *i.e.* 200°C. The zigzag edges rearrange and form armchair segments that are ±30° relative to the edge direction, while armchair edges are dominated by armchair segments even at the annealing temperature as high as 500°C. These results could be very useful in guiding future application or modification of graphene's properties based on its edges. It also provides valuable information for the device fabrication of graphene-based nanoelectronics, in which the energy gap of graphene nanostructures could be effectively controlled by tuning the concentration of each edge segments (zigzag and armchair) with thermal annealing.

**METHODS**

Graphene flakes were prepared by micromechanical cleavage on silicon wafer covered by 300 nm $SiO_2$, and single layer graphene were confirmed by Raman[39,40] and optical contrast spectra.[41-44] The Raman spectra were carried out with a WITEC CRM200 Raman system, with 532 nm excitation and 100× objective (Numerical Aperture, NA=0.95). The laser power at sample is below 1 mW to avoid possible laser heating induced sample damage.[45-47] Raman images with a spatial resolution of ~400 nm are realized with a piezo stage (step size 100 nm) and the integration time at each position is one second. The polarization-dependent Raman spectra at edges were obtained by changing the angle of incident polarization with a half wave plate. The exact position of graphene edge is determined by Raman line scan (step size 100 nm) across the edge to make sure the maximum intensity of D peak is approached.[33] Samples with angle of 30° and 150° between adjacent edges were selected so that for each sample, one edge is A-edges (strong D peak) and the other is Z-edges (weak D peak). Four different graphene flakes were annealed in vacuum ($5\times10^{-5}$ mbar) for 15 minutes at 200°C (sample b), 300°C (sample a), 400°C (sample c), and 500°C (sample d), respectively.

*Supporting Information Available:* This material is available free of charge via the Internet at http://pubs.acs.org.

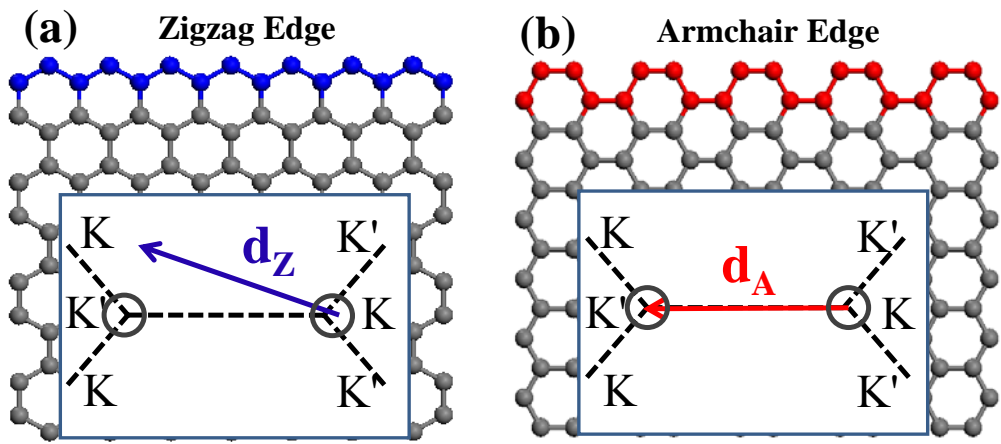

Figure 1. Schematic illustration of crystal structure of graphene edges: (a) zigzag edge; (b) armchair edge; The insets of (a) and (b) show the schematic of intervalley electron scattering in double resonance process, where zigzag/armchair segments cannot/can fulfill the scattering condition.

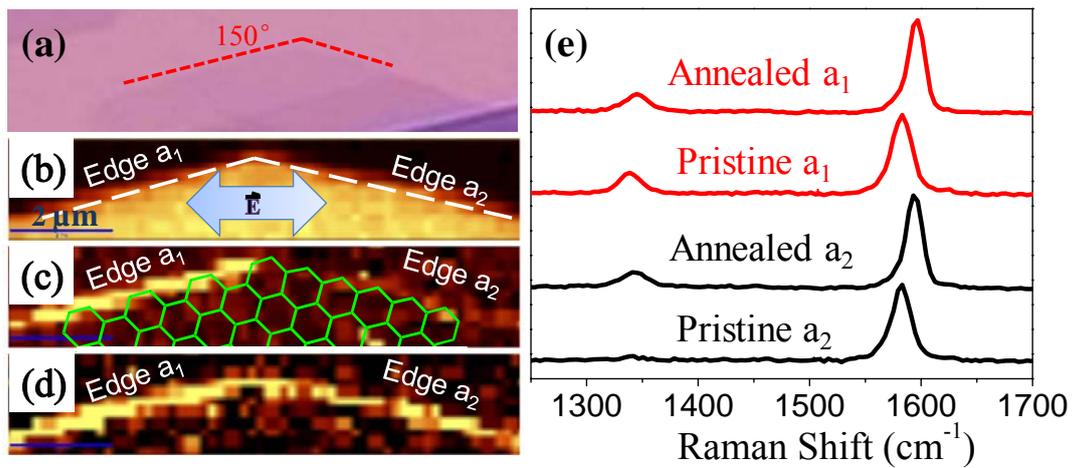

Figure 2. (a) Optical image of single layer graphene (sample a), the dashed red lines show the angle between two adjacent edges is 150°. (b) Raman image of G peak intensity of graphene (sample a) before annealing. The arrows indicate the incident laser polarization. (c) and (d) are Raman images of D peak intensity of graphene (sample a) before and after annealing at 300°C, respectively. (e) Raman spectra of pristine and annealed edges ($a_1$ and $a_2$).

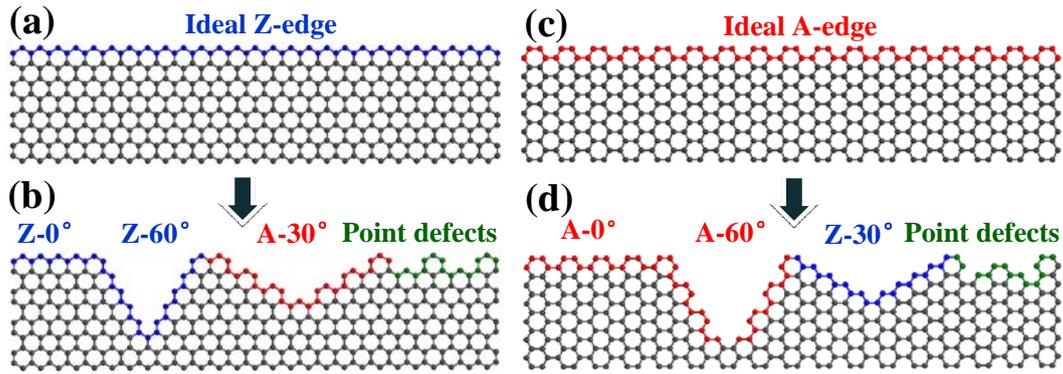

Figure 3. Schematic illustration for the edges rearrangement. (a) Ideal zigzag edge (Z-edge). (b) Annealed Z-edges contains unchanged zigzag segments (Z-0°), rearranged armchair segments at ±30° (A-30°) and zigzag segments at ±60° (Z-60°), as well as point defects (short-range defects). (c) Ideal armchair edge (A-edge). (d) Annealed A-edges contains unchanged armchair segments (A-0°), rearranged zigzag segments at ±30° (Z-30°) and armchair segments and ±60° (A-60°), as well as point defects (short-range defects). Blue atoms denote the zigzag segments, red atoms denote the armchair segments, and green atoms denote the point defects (short-range defects).

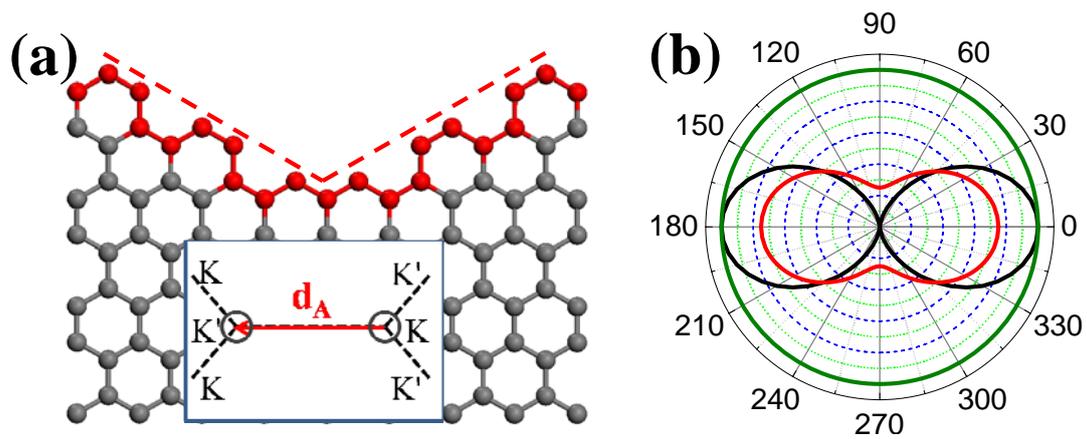

Figure 4 (a) Schematic of rearranged A-30° segments with respect to the original Z-edges direction. The inset shows that electrons scattered from such edges can fulfill the double resonance condition. (b) The normalized D peak intensity as function of the incident laser polarization for ideal A-edges (black curve) and rearranged A-30° segments (red curve) as well as point defects (green curve).

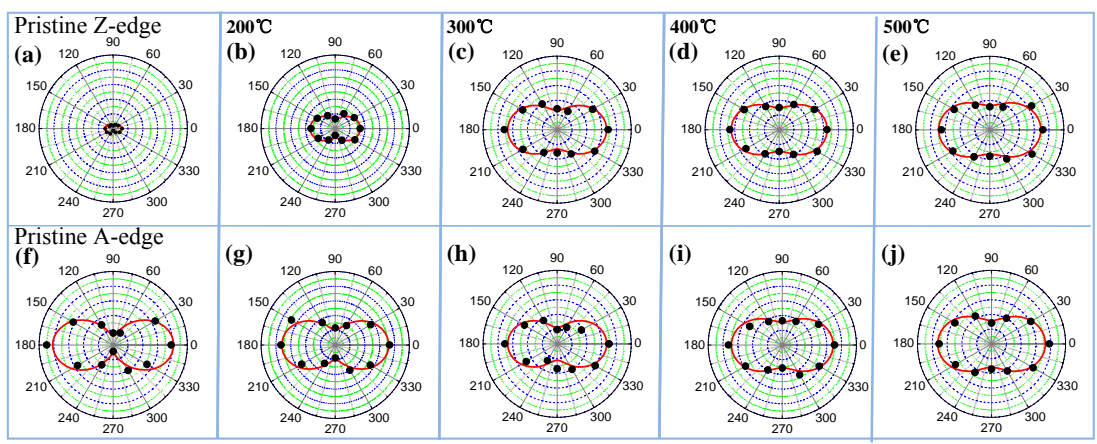

Figure 5. Polarization dependence of $I_D/I_G$, θ is the angle between laser polarization and edge direction. (a) and (f) are pristine Z-edges and A-edges, respectively. (b), (c), (d) and (e) are annealed Z-edges at 200°C, 300°C, 400°C and 500°C, respectively; (g), (h), (i) and (j) are annealed A-edges at 200°C, 300°C, 400°C and 500°C, respectively. All the polar graphs are with same scale bar, which is from 0 to 0.5.

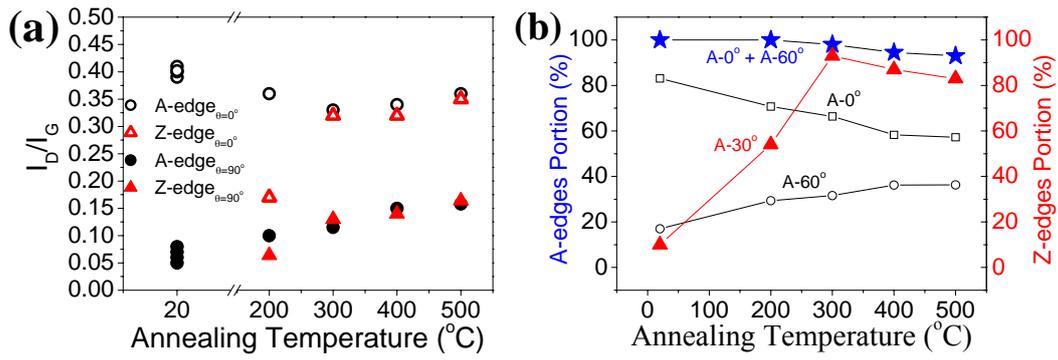

Figure 6 (a) $I_D/I_G$ of A-edges (both pristine and annealed) and Z-edges (annealed) as a function of annealing temperature. (b) Edge components evolution as a function of annealing temperature for Z-edges (red triangles for A-30°) and A-edges (black squares for A-0°; black circles for A-60° and blue stars for A-0°+A-60°), respectively.